\documentclass{article}
\usepackage{frascatiphys,graphicx}

            \def\dis{\displaystyle}
  \def\be{\begin{equation}}
          \def\ra{\rightarrow}
  \def\beq{\begin{eqnarray}}  \def\eeq{\end{eqnarray}}
  \def\be{\begin{equation}}   \def\ee{\end{equation}}

\begin{document}
\title{Bottomonium masses, decay rates and scalar charge radii}
\author {
J N Pandya        \\
{\em Department of Physics, Veer Narmad South Gujarat University,}\\ {\em Surat 395 007, INDIA.} \\
Ajay Kumar Rai        \\
{\em Applied Sciences and Humanities Department,  SVNIT,}\\ {\em Surat 395 007, INDIA.}\\
P C Vinodkumar        \\
{\em Department of Physics, Sardar Patel University,}\\ {\em Vallabh Vidyanagar 388 120, INDIA.}}
\maketitle
\baselineskip=11.6pt
\begin{abstract}
The masses of bottomonium $s$ and $p$-states, decay constants, leptonic  as well  as radiative decay widths are computed in the framework of extended harmonic confinement model without any additional parameters.
\end{abstract}
\baselineskip=14pt
\section{Bottomonia masses from ERHM}
The mass of a hadron having $p$ number of quarks in ERHM can be obtained as\cite{pcvjnp1999},
  \be
  M_N (q_1q_2.....)\ =\ \dis\sum_{i=1}^{p}\epsilon_N(q_i,p)_{conf}\ +\
  \dis\sum_{i<j=1}^{p}\epsilon_N (q_iq_j)_{coul} + \dis\sum_{i<j=1}^p \epsilon_N^J(q_i,q_j)_{SD}    \label{eq:hadronmass}\ee
where the first sum is the total confined energies of the
constituting quarks of the hadron, the second sum corresponds to
the residual colour coulomb interaction energy between the
confined quarks and the third sum is due to spin dependent
interaction.

The intrinsic energy of the quarks in a mesonic system is given by
  \be
  \epsilon_N (q_{i,2})_{conf}\ =\sqrt {(2N+3)\Omega_N (q_{i})\ +\ M_i^2\ -\ 3M_i \Omega_0(q_{i}) / (M_1+M_{2})} \label{eq:epsilon} \ee
Here $M_{i=1,2}$ represent the masses of the quark and the
antiquark constituting the meson. The coulombic part of the energy
is computed using the residual coulomb potential given
by\cite{gw1986},  $V_{coul}(q_iq_j) = k \alpha_s(\mu)/\omega_n r$,
where $\omega_n$ represents the state dependent colour dielectric
``coefficient''\cite{gw1986}. We construct the wave functions for
bottomonium by  retaining the nature of single particle wave
function but with a two particle size parameter
$\Omega_N({q_iq_j})$ instead of $\Omega_N(q)$\cite{jnppcv2001}.
Coulomb energy is computed perturbatively using the confinement
basis with two particle size parameter defined above for different
states as $ \epsilon_N(q_iq_j)_{coul}\ =\ \langle N | V_{coul} | N
\rangle$. The fitted parameters to obtain experimental ground
state mass are $m_b$ = 4637 MeV, $k = 0.19252$ and  the
confinement parameter $A = 2166$ MeV$^{3/2}$.

From the center of weight masses, the pseudoscalar and vector
mesonic masses are computed by incorporating the residual two body
chromomagnetic interaction through the spin-dependent term of the
confined one gluon exchange propagator perturbatively as $
\epsilon_N^J(q_iq_j)_{S.D.}\ =\ \langle NJ | V_{SD}|NJ \rangle $.
We consider the two body spin-hyperfine interaction of the
residual (effective) confined one gluon exchange potential
(COGEP)\cite{pcvjnp1999,vvk1992}.  The computed masses in
comparison with experimental and other theoretical model results
are given in Table \ref{tab:bbmasses}.
\section{Decay properties and scalar charge radii}
We employ radial wave functions to compute the hadronic as well as radiative decay widths of  the vector and pseudoscalar mesons of $b\bar{b}$ system based on the treatment of perturbative QCD as\cite{fec1979}. The standard Van - Royen - Weisskopf formula has been used without radiative correction term for computing leptonic decay widths\cite{qr1979}. The computed leptonic decay widths are tabulated in Table \ref{tab:leptdcaywdth} alongwith other theoretical as well as experimental values.

The Van Royen - Weisskopf formula used for the meson decay constants is obtained in the two-component spinor limit\cite{vrw1967}. $f_P$ and $f_V$ are related to the ground state radial wave function $R_{1S}(0)$ at the origin, by the VR-W formula modified for the colour as,
  $ f_{P/V}^2\ =(3/\pi M_{P/V}) |R_{1S}(0)|^2$,  where $M_{P/V}$ is the ground state mass of the pseudoscalar/vector meson. In the present computations, $f_P$ = 711 MeV which is closer to the experimental value of 710 $\pm$ 15 MeV, while the other results are 660 \cite{qr1979}, 772 \cite{cetal2004} and 812 \cite{zw2005}.

The scalar charge radii and M1 transitions  of the bottomonia in a given eigenstate are obtained using $ \langle r_{nS}^2 \rangle^{1/2}\ =\ \left [\int_0^{\infty}  |R_{nS}^{h}(r)|^2\ r^2\ r^2 dr \right ]^{1/2} $,
$ \Gamma (V\ra P\gamma) = \frac{16}{3}\alpha e_q^2\frac{k_\gamma^3}{M_V^2} $ where $k_\gamma = (M_V^2 - M_P^2)/2M_V$ is the energy of the emitted photon. The computed values of M1 transitions are shown in Table \ref{tab:m1trans}. The scalar charge radii (in fm) of $s$-wave bottomonia are 0.1854 (2.2338), 0.3997 (1.6325), 0.7070 (1.0890) and 1.1649 (0.9623) respectively from 1S through 4S states, where the values in the brackets are wave functions (in GeV$^{3/2}$).

  \begin{table}\centering
  \caption{Masses (in MeV/c$^2$) of the bottomonium system}\label{tab:bbmasses}
  \begin{tabular}{ccccccccc}
  \hline\hline
  State  &  Present  & \cite{pdg2006} &\cite{rac2006}  & \cite{sfr2007}$^{p}$ & \cite{sfr2007}$^{np}$ & \cite{de2003}  & \cite{avs2003}  \\
  \hline
  $\eta_b(1^1S_0)$ & 9425   & 9300 $\pm$ 23 & 9457  & 9414  & 9421 & 9400  &  9300 \\
  $\eta_b(2^1S_0)$ & 10012  &  --  & 10018 & 9999  & 10004 & 9993  & 9974     \\
  $\eta_b(3^1S_0)$ & 10319  &  --  & 10380 & 10345 & 10350 & 10328 & 10333    \\
  $\eta_b(4^1S_0)$ & 10572  &  --  & 10721 & 10623 & 10632 & --    & --      \\
  $\eta_b(5^1S_0)$ & 10752  &  --  & 11059 &  --   & --    & --    & --     \\
\hline
  $\Upsilon(1^3S_1)$ & 9461   & 9460   & 9460  & 9461  & 9460  & 9460   & 9460  \\
  $\Upsilon(2^3S_1)$ & 10027  & 10023  & 10023 & 10023 & 10024 & 10023  & 10023   \\
  $\Upsilon(3^3S_1)$ & 10329  & 10355  & 10385 & 10364 & 10366 & 10355  & 10381  \\
  $\Upsilon(4^3S_1)$ & 10574  & 10579  & 10727 & 10643 & 10643 &  --    & 10787 \\
  $\Upsilon(5^3S_1)$ & 10753  & 10865  & 11065 & --    & --    &  --    & 11278  \\
\hline
  $\chi_{b0}(1^3P_0)$ & 9839  & 9859  & 9894 & 9861 & 9860 & 9863  & 9865  \\
  $\chi_{b1}(1^3P_1)$ & 9873  & 9893  & 9941 & 9891 & 9892 & 9892 & 9895  \\
  $\chi_{b2}(1^3P_2)$ & 9941  & 9912  & 9983 & 9912 & 9910 & 9913 & 9919  \\
  $h_{b1}(1^1P_1)$    & 9907  &  --   & 9955 & 9900 & 9900 & 9901   & 9894  \\
\hline
  $\chi_{b0}(2^3P_0)$ & 10197  & 10232  & 10234 & 10231 & 10231 & 10234  & 10238  \\
  $\chi_{b1}(2^3P_1)$ & 10207  & 10255  & 10283 & 10255 & 10258 & 10255 & 10264  \\
  $\chi_{b2}(2^3P_2)$ & 10227  & 10268  & 10326 & 10272 & 10271 & 10268 & 10283 \\
  $h_{b2}(2^1P_1)$    & 10217  &  --    & 10296 & 10262 & 10263 & 10261  & 10260 \\
  \hline\hline
\end{tabular}\\
$p$ = perturbative and $np$ = nonperturbative computations in Tables \ref{tab:bbmasses} \& \ref{tab:leptdcaywdth}
\end{table}

\begin{table} \centering
\caption{Leptonic decay widths (in keV) of
$\Upsilon(n^3S_1)$}\label{tab:leptdcaywdth}
\begin{tabular}{cccccccc}
  \hline\hline
  State & Present & \cite{pdg2006} & \cite{rac2006} & \cite{sfr2007}$^{p}$ & \cite{sfr2007}$^{np}$& \cite{smi2006} & \cite{vva2007}  \\
  \hline
  $\Upsilon(1^3S_1)$ & 1.320 & 1.340 $\pm$ 0.018 & -- & 5.30 & 1.73 & 1.45 $\pm$ 0.07 & 1.314 \\
  $\Upsilon(2^3S_1)$ & 0.628 & 0.612 $\pm$ 0.011 & 0.426 & 2.95 & 1.04 & 0.52 $\pm$ 0.02 & 0.576 \\
  $\Upsilon(3^3S_1)$ & 0.263 & 0.443 $\pm$ 0.008 & 0.356 & 2.17 & 0.81 & 0.35 $\pm$ 0.02 & 0.476 \\
  $\Upsilon(4^3S_1)$ & 0.104 & 0.272 $\pm$ 0.029 & 0.335 & 1.67 & 0.72 & -- & 0.248 \\
  $\Upsilon(5^3S_1)$ & 0.040 & 0.310 $\pm$ 0.070 & 0.311 & -- & -- & -- & 0.310 \\
  \hline\hline
\end{tabular}\\
\end{table}
  \begin{table}\centering
  \caption{Radiative M1 transitions of bottomonia (eV)}\label{tab:m1trans}
  \begin{tabular}{ccccccc}
  \hline\hline
  Transition  &  Present &\cite{sfr2007}&\cite{de2003}& \cite{vva2007} &\cite{tal2002}
  &\cite{wa2004}\\
  \hline
  $1^3S_1\ra 1^1S_0$ & 2.242 (36) & 4.0 & 5.8 (60)& 9.2 & 7.7 (59)  &8.95\\
  $2^3S_1\ra 2^1S_0$ & 0.145 (15) & 0.5 & 1.40 (33)& 0.6 & 0.53 (25) &1.51 \\
  $3^3S_1\ra 3^1S_0$ & 0.041 (10) & -- & 0.80 (27)& -- & 0.13 (16) &0.826 \\
  \hline\hline
  \end{tabular}\\
  The values in the parentheses are the energy of emitted photons in MeV.
  \end{table}

\baselineskip=11pt
\begin{thebibliography}{99}
  \bibitem{pcvjnp1999} P C Vinodkumar {\em et al}; Eur. Phys. Jnl. {\bf A4}, 83 (1999)
  \bibitem{gw1986} K Gottfried, V F Weisskopf, Concepts of Particle Physics; 397 (1986)
  \bibitem{jnppcv2001} J N Pandya, P C Vinodkumar, Pramana J. Phys {\bf 57}, 821 (2001)
  \bibitem{vvk1992} P C Vinodkumar {\em et al}; Pramana Jnl. of Phys. {\bf 39}, 47 (1992)
  \bibitem{fec1979} Close F E, Quarks and Partons (1979)
  \bibitem{qr1979} C Quigg, J L Rosner, Phys. Reports {\bf 56}, 222 (1979)
  \bibitem{vrw1967} R Van Royen, V F Weisskopf, {\em Nuovo Cimento} {\bf A50}, 617 (1967)
  \bibitem{cetal2004} Cvetic {\em et al}, Phys. Lett. {\bf B596}, 84 (2004)
  \bibitem{pdg2006} Y M Yao {\em et al} [Particle Data Group]; J. Phys. {\bf G33},1 (2006)
  \bibitem{zw2005} Z Wang et al, Phys. Lett. {\bf B 615}, 79 (2005)
  \bibitem{rac2006} R A Coimbra, O Oliveira; hep-ph/0610142 (2006)
  \bibitem{sfr2007} S F Radford, W W Repko; Phys. Rev. {\bf D75},  074031 (2007)
  \bibitem{de2003} D Ebert, R N Faustov, V O Galkin; Phys. Rev. {\bf D 67}, 014027 (2003)
  \bibitem{avs2003} A V Shoulgin, G M Vereshkov, O V Lutchenko; J. Phys. {\bf G29}, 1245 (2003)
  \bibitem{smi2006} S M Ikhdair, R Sever; Int. J. Mod. Phys. {\bf A 21}, 3989 (2006)
  \bibitem{vva2007} V V Anisovich {\em et al}; Phys. Atom. Nucl. {\bf 70}, 63 (2007)
  \bibitem{tal2002} T A L\"ahde; Nucl.Phys. {\bf A714}, 183 (2003)
  \bibitem{wa2004} World average, N Brambilla et al; hep-ph/0412158v2
\end{thebibliography}
\end{document}